\documentclass[pra,twocolumn,showpacs,amsfonts,amsmath,floatfix]{revtex4}

\usepackage{graphicx}
\usepackage{epsfig}
\usepackage{epstopdf}
\usepackage[usenames,dvipshot noiseames]{xcolor}

\newcommand{\lt}{\left(}
\newcommand{\rt}{\right)}
\newcommand{\lqu}{\left[}
\newcommand{\rqu}{\right]}
\newcommand{\lgr}{\left\{}
\newcommand{\rgr}{\right\}}

\newcommand{\be}{\begin{equation}}
\newcommand{\ee}{\end{equation}}
\newcommand{\ba}{\begin{eqnarray}}
\newcommand{\ea}{\end{eqnarray}}
\newcommand{\fr}{\frac}
\newcommand{\nn}{\nonumber}
\newcommand{\se}{\section}

\begin{document}

\title{Optimization of networks for measurement based quantum computation}

\author{G. Ferrini,  J. Roslund, F. Arzani, Y. Cai, C. Fabre and N. Treps} 
\affiliation{Laboratoire Kastler Brossel, UPMC Univ. Paris 6, ENS, CNRS; 4 place Jussieu, 75252 Paris, France} 

\begin{abstract}

In this work we provide a recipe for mitigating the effects of finite squeezing, which affect the production of cluster states and the result of a measurement based quantum computation in the continuous variable regime. 
To this aim, we derive a compact expression for the unitary matrix which describes the linear optics network that generates a certain cluster state from independently squeezed inputs. We show that this possesses tunable degrees of freedom, which can be exploited to minimize the noise effects.
These strategies are readily implementable by several experimental groups. 

\end{abstract}
\maketitle

\section{Introduction}

Continuous Variable (CV) quantum computing in the measurement based approach~\cite{Briegel2001,Raussendorf2001one} has gained significant interest following recent experiments in which large cluster states have been constructed with time~\cite{yokoyama2013optical} or frequency~\cite{Chen2014} encoding. The ability to perform a quantum computation (QC) on a resource state that is consumed in time opens the possibility of scaling the computation to large mode numbers. 
One of the difficulties associated with CV measurement based quantum computing (MBQC) is that finite squeezing engenders errors that propagate through the computation and contribute extra noise to the result~\cite{Nielsen2006,Gu2009}. Recently, however, a fault tolerant CV MBQC protocol was proposed~\cite{Menicucci2014} that utilizes extra ancillary modes and necessitates the squeezing of each input mode to be above a finite, albeit demanding, threshold.
This result is encouraging and supports the positive outlook for CV MBQC. However, a need still exists for QC protocols that minimize errors as much as possible for a given degree of squeezing and do so without introducing additional resources. 

In the traditional MBQC approach, the employed resource state is a cluster state. 
This is a highly entangled state for which the variances of specific quadrature combinations, called the nullifiers, tend to zero in the limit of infinite squeezing. These nullifiers are
expressed by
\be
\label{eq:quadratures_cluster}
\hat \delta_i \equiv \lt \hat{p}_i^{C} - \sum_l V_{il} \hat{x}_l^{C} \rt   \hspace{0.25cm} \forall  \hspace{0.1cm}  i = 1,..., N,
\ee
where $V$ is the adjacency matrix associated to the cluster state and $\hat{x}_i^{C}, \hat{p}_i^{C}$ are the amplitude and phase quadratures associated with each mode $i$ of the cluster state~\cite{Nielsen2006, Gu2009}. 
A $N$-mode cluster state in CV can be constructed by sending $N$ squeezed modes into a suitably chosen linear optical network~\cite{vanLoock2007, vanLoock2008, Su07, Furusawa2011, Su12}.

In this work we provide an optimization strategy for quantum computation that diminishes errors arising from finite squeezing in the modes used to create the ancillary cluster state. In particular, if the squeezing degree is not equivalent for all of the input modes, this approach optimally redistributes the available correlations among the transformed modes. The ability to employ optimization strategies is shown to result from tunable degrees of freedom contained within the unitary matrix that fabricates cluster states from a set of squeezed states. 

The linear optical network associated with a given unitary matrix is most commonly implemented by a series of optical elements, such as beam splitters and phase shifters, when the input squeezed modes are each embedded in a different spatial mode. The layout of the optical network is determined by a specific decomposition procedure~\cite{vanLoock2007}. Our optimization method could therefore be readily implementable in any experimental group, as it provides the optimal unitary matrix that can be decomposed in the corresponding practical series of optical elements.

After setting the problem in Sec.\ref{se:problem}, we derive a compact analytical expression in Sec.\ref{se:Derivation} for the unitary matrix which allows the generation of a cluster state starting from squeezed states; this expression possesses tunable degrees of freedom. Then, we show that these degrees of freedom can be exploited to reduce noise in a cluster state measurement (Sec.\ref{se:opt-clu}) and also in the result of a quantum computation (Sec.\ref{se:opt-mbqc}). We conclude in Sec.\ref{se:conclusions}.

\section{Problem setting}
\label{se:problem}

Consider a general linear transformation, corresponding to a specific unitary matrix $U$, which acts on an initial set of squeezed modes $\vec{a}^{\, \text{squ}} = (\hat{a}^{\text{squ}}_1,...,\hat{a}^{\text{squ}}_N)$:
\be
\label{eqtrasf-y0}
\vec{a} \, ' =  U \vec{a}^{\, \text{squ}},
\ee
where $\vec{a} \, ' = (\hat{a}^{'}_1,...,\hat{a}^{'}_N) $ indicates the set of annihilation operators associated with the modes at the output of the transformation $U$.
A given task, such as the creation of cluster states or the execution of a QC, is accomplished through judicious selection of the $U$ matrix. This matrix is not unique, however, and several internal degrees of freedom may be exploited  to minimize errors associated with the task. 

As an example, in the case of cluster state creation with finitely squeezed inputs, $U$ may be chosen so as to minimize the mean of the nullifier variances: 
\be
\label{eq:f1}
f_1 = \fr{1}{N} \sum_{i = 1}^N \fr{\Delta^2 \delta_i}{\Delta^2  \delta_i^0},
\ee
where  $\lgr \Delta^2 \hat \delta_i \rgr$ are the variances of the nullifiers defined in Eq.(\ref{eq:quadratures_cluster}), and $\lgr \Delta^2 \hat \delta_i^0 \rgr$ are the shot noise variances, defined as the nullifier variances for vacua inputs. This choice of function $f_1$ is not unique, and other analogous functions may be defined \footnote{We obtain similar results by defining $\tilde{f}_1 = \text{Mean}\lqu \Delta^2 \hat \delta_i - \Delta^2 \hat \delta^0_i \rqu$.}. 

In the context of MBQC, the result is encoded in the output mode, which is affected by the excess noise incurred by finite squeezing. This can be reduced by minimizing
~\footnote{
Other choices are once again possible, e.g. one may minimize the noise of a single quadrature only. This may be useful depending upon the specific output mode readout.}
\be
\label{eq:fitness3}
f_2 = 
\Delta^2 {\hat{x}_{\text{extra}} } +  \Delta^2 {\hat{p}_{\text{extra}} }.
\ee  
As will be discussed later, those noise terms can be related to the nullifiers of the ancillary cluster used for the computation.

\se{Derivation of a general expression for the cluster unitary network} 
\label{se:Derivation}

It has been shown in Ref.~\cite{vanLoock2007} that a cluster state with an adjacency matrix $V$ is obtained by applying a unitary matrix $U = U_V = X + i Y$ to a set of input squeezed modes as in Eq.(\ref{eqtrasf-y0}). This unitary matrix must satisfy the condition 
\be
\label{eq:x_and_y}
Y - V X  = 0. 
\ee
Then, the graph of the state expressed by (\ref{eqtrasf-y0}) coincides with $V$ in the limit of infinite squeezing (see Appendix~\ref{app:A}).

We now seek an explicit and compact analytical expression for the unitary transformation satisfying Eq.(\ref{eq:x_and_y}).
The unitary transformation of Eq.(\ref{eqtrasf-y0}) on the annihilation operators induces a transformation on the quadratures of each mode which is expressed by the symplectic matrix 
\be
\label{eq:boh199}
\left( \begin{array}{cccccccc}
\vec{x}^{\, C} \\
\vec{p}^{\, C} \\
\end{array} \right) =
\left( \begin{array}{cccccccc}
X & - Y \\
Y &  X \\
\end{array} \right)
\left( \begin{array}{cccccccc}
\vec{x}^{\, \text{squ}} \\
\vec{p}^{\, \text{squ}} \\
\end{array} \right)  \\
\ee
where $X$ and $Y$ are defined as in Eq.(\ref{eq:x_and_y}), and where $(\vec{x}, \vec{p})^T = (\hat{x}_1, ..., \hat{x}_{N},\hat{p}_{1}, ..., \hat{p}_{N})^T$. 
In order to be symplectic, the matrix of Eq.(\ref{eq:boh199}) must satisfy~\cite{dutta1995real}
\ba
&& X X^T + Y Y^T = \mathcal{I} \label{eq:1} \\
&& X Y^T = Y X^T  \label{eq:2}
\ea  
where $\mathcal{I}$ is the identity matrix.
With the use of Eq.(\ref{eq:x_and_y}), Eq.(\ref{eq:1}) and (\ref{eq:2}) one obtains 
\be
\label{eq:symme-eq}
X X^T = (V^2 + \mathcal{I})^{-1}.
\ee
Hence, the symmetric solution $X_s = X_s^T$ is simply given by
\be
\label{eq:sol}
X_s = (V^2 + \mathcal{I})^{-1/2}
\ee
from which, using Eq.(\ref{eq:x_and_y}), one obtains  that $Y_s = V  (V^2 + \mathcal{I})^{-1/2}$ and hence
\be
\label{eq:sol_symm}
{U_V}_s =(1 + i V)  (V^2 + \mathcal{I})^{-1/2}.
\ee
Eq.(\ref{eq:sol_symm}) represents the symmetric solution ${U_V}_s = {U_V}_s^T$ for the linear network we were seeking.

Furthermore, notice that if $X_s$ and $Y_s$ satisfy Eq.(\ref{eq:x_and_y}), then $X = X_s \mathcal{O}$ and $Y = Y_s \mathcal{O}$  are also a solution for any real orthogonal matrix $\mathcal{O}$ due to the symmetry of Eq.(\ref{eq:symme-eq}), where $\mathcal{O} \mathcal{O}^T = \mathcal{I}$. Hence, we obtain the general solution for the unitary matrix yielding a cluster state with graph $V$, which is provided by 
\be
\label{eq:sol_all}
{U_V} (\vec{\theta})  =(\mathcal{I} + i V)  (V^2 + \mathcal{I})^{-1/2} \mathcal{O} (\vec{\theta}),
\ee
where we have discarded the subscript ``s", and where we have rendered explicit the parameterization of $\mathcal{O}$ in terms of angular variables $\vec{\theta}$. These are $N (N-1)/2$ degrees of freedom, and can be chosen e.g. as Euler or Tait-Bryan angles.

\se{Optimization of cluster states} 
\label{se:opt-clu}

The real orthogonal matrix $\mathcal{O}(\vec{\theta})$ appearing in Eq.(\ref{eq:sol_all}) may be freely chosen to mitigate the effects of finite squeezing on cluster state preparation and MBQC. Selection of an optimal matrix is achieved by employing an evolutionary strategy~\cite{roslund2009accelerated}, which is particularly suitable for high-dimensional parameter searches. In practice, we search for a $\vec{\theta}$ that minimizes a fitness function, such as the ones previously described. 
Having discovered an optimal orthogonal matrix, the unitary matrix implementing the desired cluster state is fully specified by Eq.(\ref{eq:sol_all}). 

If the squeezing levels for the input modes are uniform, all of the unitaries described by Eq.(\ref{eq:sol_all}) are equivalent, in that each nullifier of the resulting cluster state possesses the same level of noise. 
For nonuniform squeezing levels, however, one may search for a unitary matrix $U_V$ that redistributes the available squeezing among the modes in a manner that optimizes some desired property, e.g., $f_1$ in  Eq.(\ref{eq:f1})~\footnote{An optimization following this spirit is performed analytically in Ref.~\cite{menicucci2011graphical} in a different context, with respect to local phase shifts only.}. 

\begin{figure} 
\begin{minipage}{.98\columnwidth}
\includegraphics*[width=0.75\columnwidth]{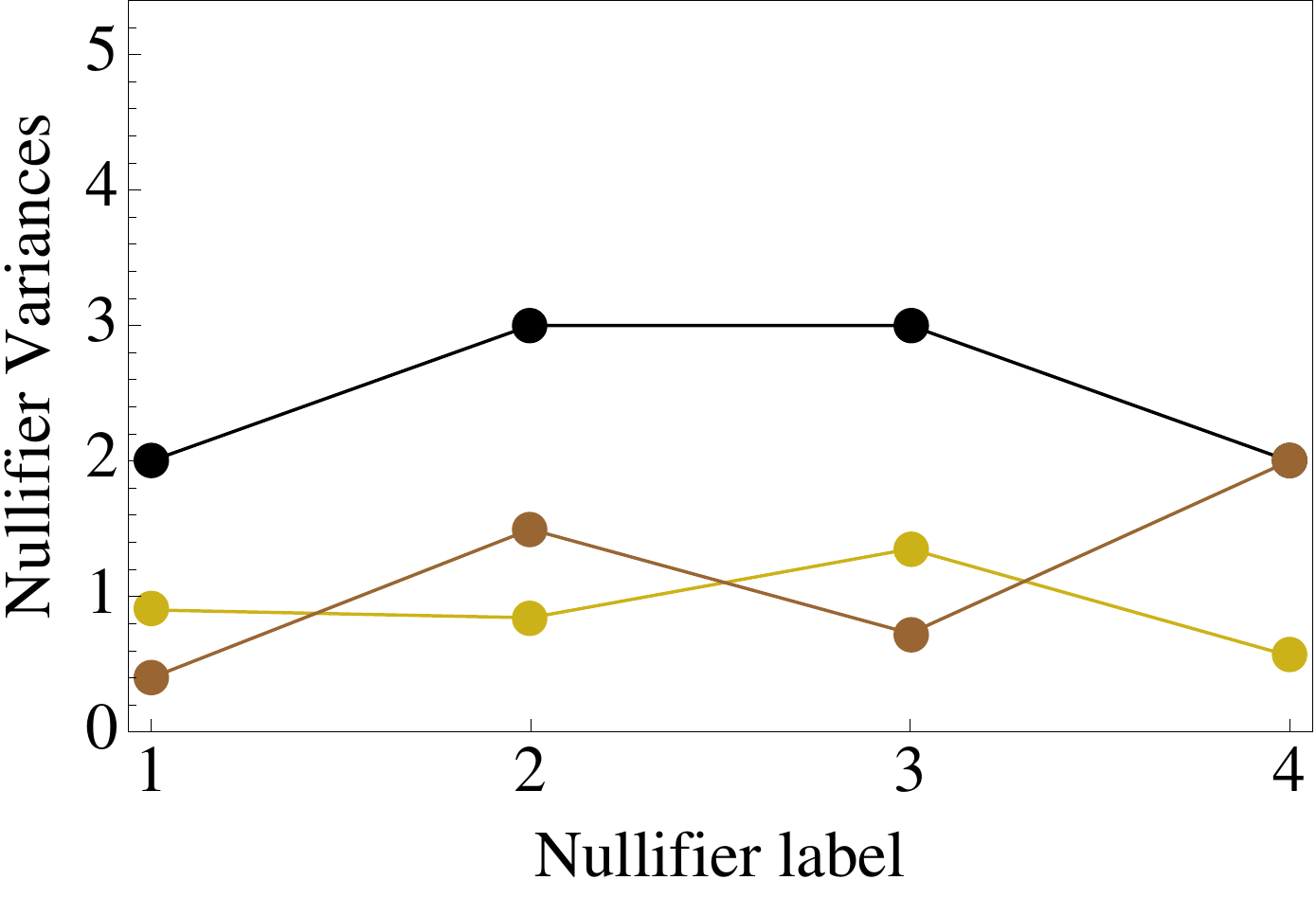}
\end{minipage}
\begin{minipage}{.98\columnwidth}
\includegraphics*[width=0.85\columnwidth]{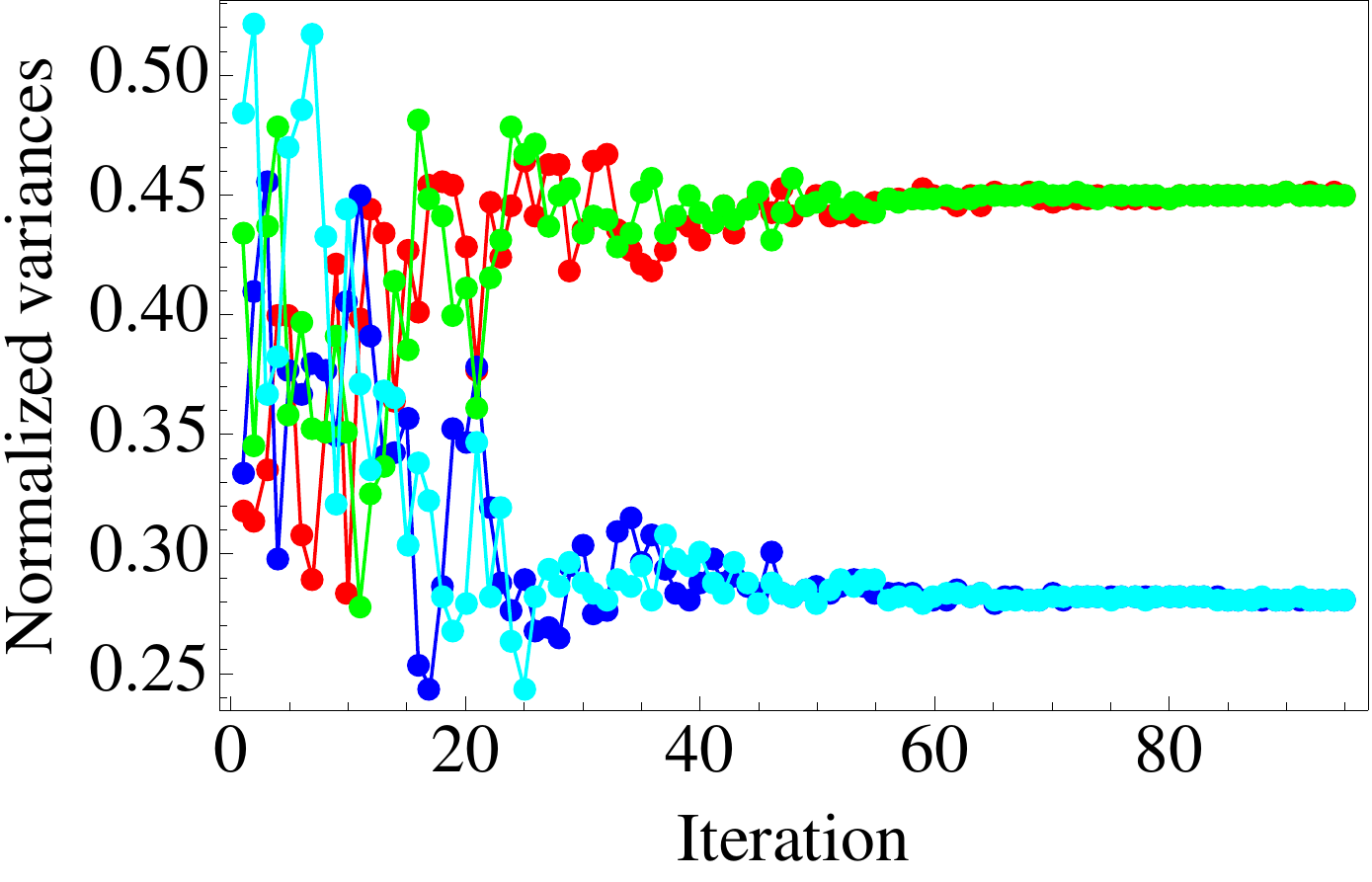}
\end{minipage}
\caption{(Color Online) Top: linear cluster state nullifier variances $\lgr \Delta^2 \hat \delta_i \rgr$ for a non-optimized network (brown / gray), for the optimized network (yellow / light gray), as compared to the shot noise levels (black). Bottom: normalized nullifiers variances  $\lgr \Delta^2 \hat \delta_i / \Delta^2 \hat \delta_i^0 \rgr$ as a function of the iteration (``generation" in the language of evolutionary strategies) of the algorithm. Upper lines: The red (dark) line is $\Delta^2 \hat \delta_1 / \Delta^2 \hat \delta_1^0$, the green (light) $\Delta^2 \hat \delta_3 / \Delta^2 \hat \delta_3^0$. Lower lines: The blue (dark)  is $\Delta^2 \hat \delta_2 / \Delta^2 \hat \delta_2^0$, the cyan (light) line is  $\Delta^2 \hat \delta_4 / \Delta^2 \hat \delta_4^0$.}
\label{fig:optimization}
\end{figure}

As an example, we consider the fabrication of a 4-mode linear cluster state. This state has been shown to be a universal resource for gaussian single-mode quantum computation~\cite{Ukai2010b}.
The input modes used to build the cluster state are assumed to possess realistic squeezing levels, such as those seen in the four-mode multimode state of Ref.~\cite{Roslund_13b}. Specifically, the squeezed quadrature variances relative to shot noise level are taken as $-7$dB, $-6$dB, $-4$dB, and $0$dB. The use of these modes to fashion a cluster state from the $U_{V}$ defined in Ref.\cite{vanLoock2008} yields the nullifier variances reported in Table~\ref{table-cluster} (with the convention $\Delta^2_{vac} = 1$), where the final nullifier lies at the shot noise level. Yet, by optimizing the angles $\vec{\theta}$ in Eq.(\ref{eq:sol_all}) to minimize $f_1$, all of the nullifier variances are lowered below the shot noise level and $f_1$  is reduced by $25 \%$. Consequently, the optimization successfully reduces the residual error induced by finite squeezing. Importantly, this procedure may be applied to a cluster state of arbitrary dimension as well as to the case of non-pure states. 
The absolute values of each optimized nullifiers, as compared to the nullifiers obtained with the non-optimized network defined in Ref.\cite{vanLoock2008},  as well as the shot noise values, are visualized in the upper panel of Fig.\ref{fig:optimization}.
The lower panel of Fig.\ref{fig:optimization} instead represents how each nullifier converges to the optimized solution, highlighting how the fluctuations are re-distributed among the modes.
\begin{table}

\centering

\begin{tabular}{ l | c | r | } 
                  
  {} & Nullifier variances $\{ \fr{\Delta^2 \delta_i}{{\Delta^2 \delta_i}_0} \}$ & $f_1$ \\   \hline     
  Network $U_V$ from Ref.\cite{vanLoock2008}  &  $\{0.20, 0.50, 0.24, 1.00\}$ &  0.49 \\   \hline     
  Optimized network $U_V^{\text{best}}$ & \{0.45, 0.28, 0.45, 0.28\} & 0.37 \\     \hline  

\end{tabular}
\caption{Comparison of nullifier variances for a 4-node linear cluster state with and without optimization of the unitary transformation $U$.  ${\Delta^2 \delta_i}_0$ are the shot noise (shot noise) levels, which are defined as the nullifier variances for vacua inputs.}

\label{table-cluster}

\end{table}

\se{MBQC error reduction}
\label{se:opt-mbqc}

The ability to minimize cluster state nullifiers may be directly translated to the task of reducing errors of a MBQC in the cluster-based approach~\cite{vanLoock2007}. 
Consider a Gaussian single-mode MBQC, where the state that has to be processed is encoded by a single optical mode, and we want to implement a Gaussian transformation on this state. This input state is first ``attached" to a linear cluster state via teleportation: e.g., a beam-splitter interaction is applied between the input mode and the first mode of the cluster, and a measurement of the quadratures of these two modes follows. 
A properly performed quadrature measurement effectively projects the unalterated input state on the second mode of the cluster state, apart from displacement operators which depend on the outcomes of the measurement, and which can be accounted for at the end of the calculation. Then, the second and, in general, third modes of the cluster state are measured on suitable quadratures, which depend on the computation that we want to perform and are determined with the recipe of Ref.~\cite{Ukai2010b}. These quadrature measurements on each mode can be equivalently described as a suitable rotation matrix $D_{\text{meas}}$, followed by a measurement in the same quadrature on all the modes, say, $\hat p$. In other words, measuring $\hat p$ on the modes $in$ and $1$ after having rotated them by $\pi/2$ corresponds to a measurement of $\hat{x}_{\text{in}}$ and $\hat{x}_1$. 
Note that since all the measurements performed here are Gaussian, they can be done simultaneously - no adaptation of the measurement basis is effectively needed~\cite{Gu2009}.

Hence, in summary, the unitary in Eq.(\ref{eqtrasf-y0}) results from the product of three matrices: the unitary $U_V$ constructing the cluster from the input squeezed modes, the beamsplitter interaction between the input state and a single mode of the cluster, and a proper diagonal matrix specifying each mode's measurement quadrature:
\be
\label{eq:q_comp}
U = U_{\text{comp}} = D_{\text{meas}} U_{\text{BS}} U_{V}(\vec{\theta}).
\ee

As an example, we consider a Fourier transformation of the input state, i.e. $\left(  \begin{array}{cccccccc} 
\hat{x}_{\text{out}} \\
\hat{p}_{\text{out}} \\ \end{array}
\right) =
\left(  \begin{array}{cccccccc}
0 & -1 \\
1 & 0 \\
\end{array} \right)
\left( \begin{array}{cccccccc}
\hat x_{\text{in}} \\
\hat p_{\text{in}} \\
\end{array} \right) =
\left(  \begin{array}{cccccccc}
- \hat p_{\text{in}} \\
\hat x_{\text{in}} \\
\end{array} \right)$. 
The measurement matrix $D_{\text{meas}}$ along with the $U_{V}$ necessary to implement this QC by a three-mode cluster state are reported in Ref.~\cite{ferrini2013compact} (see also Appendix \ref{app:B}). The calculation of the output mode containing the computation result follows from Refs.~\cite{vanLoock2007,Furusawa_Book} and yields (see Appendix \ref{app:B})
\ba
\label{eq:teleport_result4}
\hat{x}_{\text{out}}  &=& \hat{x}'_3 = -\hat{p}_{\text{in}} + {p}'_2 - \sqrt{2} {p}'_1 - \hat{\delta}_2  \\
\hat{p}_{\text{out}} &=& \hat{p}'_3 =  \hat{x}_{\text{in}}   - \sqrt{2} p'_{\text{in}}  - \hat{\delta}_1 + \hat{\delta}_3  \nn,
\ea
where $\hat{\delta}_{i}$ are the previously defined nullifiers depending upon the chosen realization of $U_{V}(\vec{\theta})$, and where $p_i$ are real numbers, given by the results of the quadrature measurements on modes $in, 1, 2$. 
By optimizing $\mathcal O(\vec{\theta})$, we have already shown that the nullifier variances are minimized, thereby reducing the error of the QC.

As a demonstration of the optimization scheme in the context of MBQC, the squeezing distribution of Ref.~\cite{Roslund_13b} is again considered where the fourth (minimally squeezed) mode serves as the input mode.
In the absence of any optimization (i.e., $\mathcal{O}(\vec{\theta})  = \mathcal{I}$ and $U_{V}(\vec{\theta})$ is given in Ref.~\cite{ferrini2013compact}), the excess noise quadratures $\Delta^2 {\hat{x}_{\text{extra}} } = \Delta^2 \hat{\delta}_2 $ and $\Delta^2 {\hat{p}_{\text{extra}} } = \Delta^2(- \hat{\delta}_1 + \hat{\delta}_3 )$ in Eq.(\ref{eq:teleport_result4}) are detailed in Table~\ref{table-QC}
\footnote{As opposed to the output modes of Eq.(\ref{eq:teleport_result2}), the extra noise modes are not constrained by the uncertainty principle $\Delta^2 {\hat{x} } \Delta^2 {\hat{p }} \geq 1$ as they are added contributions, and can display $\Delta^2 {\hat{x}_{\text{extra}} } +\Delta^2 {\hat{p}_{\text{extra}} }  = 0$ in the limit of high squeezing. The shot noise limit yields $\Delta^2 {\hat{x}_{\text{extra}} } = 3$ and $\Delta^2 {\hat{p}_{\text{extra}} } = 2$, i.e. $\Delta^2 {\hat{x}_{\text{extra}} } + \Delta^2 {\hat{p}_{\text{extra}} } = 5$.}. 
By optimizing $f_2$ over $\vec{\theta}$, it is possible to suppress the total noise by $\sim 35 \%$ (see Table~\ref{table-QC}). It is important to note that the obtained noise reduction is specific to the distribution of input squeezing values. Nonetheless, these results readily generalize to more complicated clusters, including the support of multimode operations. 

\begin{table}

\centering

\begin{tabular}{ l | c |  c |   c | r | }

   {} & $\Delta^2 {\hat{x}_{\text{extra}} } $ & $\Delta^2 {\hat{p}_{\text{extra}} } $ &  $f_2$   \\
    \hline        
  $U_{\text{comp}}$ from Ref.\cite{ferrini2013compact}   & 1.20 &  0.48 &    1.70 \\   \hline  
  Optimized $U_{\text{comp}}^{\text{best}}$ & 0.60 & 0.50 & 1.10 \\ \hline  

\end{tabular}

\caption{Comparison of a QC's excess noise with and without optimization of the unitary transformation $U$.}

\label{table-QC}

\end{table}

\se{Conclusions}
\label{se:conclusions}

In conclusion, we have demonstrated the use of optimization strategies to mitigate noise in MBQC that arises from finite squeezing. Within the traditional cluster-based framework, the transformation between the input squeezed modes and the desired network possesses several tunable degrees of freedom, which may be optimized in order to reduce the state's excess noise.

We stress that any experimental group investigating MBQC on cluster states generated by a linear optical network $U_V$ may employ these optimization strategies to mitigate errors associated with finite and nonuniform squeezing of the input modes. For instance, Refs.~\cite{Cai2014, Wang2014} detail the production of squeezed states from four-wave mixing in a hot rubidium vapour. The squeezed states produced in this vapor display a non-uniform squeezing distribution, which is ideally suited for application of our method.

This optimization strategy has already been successfully employed in the context of an experiment led by several of the authors. In particular, given an optimal choice of the cluster unitary matrix, cluster structures are observable within the quantum state generated by a synchronously pumped optical parametric oscillator~\cite{Roslund_13b}.

Similar protocols for noise reduction have likewise been discussed in the context of dual rail encoding~\cite{Alexander2014}. All of these schemes still remain within the domain of Gaussian transformations, and the inclusion of a non-Gaussian operation will prove necessary in order to provide an advantage with respect to classical computing~\cite{Mari2012}.

\se{Acknowledgments}

This work is supported by the European Research Council starting grant Frecquam and the French National Research Agency project Comb. C.F. is a member of the Institut Universitaire de France. J.R. acknowledges support from the European Commission through Marie Curie Actions, and Y.C. recognizes the China Scholarship Council. 

\appendix

\se{Derivation of Eq.(17) in Ref.~\cite{vanLoock2007}}
\label{app:A}

Here we derive Eq.(\ref{eq:x_and_y}), which coincides with Eq.(17) in Ref.~\cite{vanLoock2007}, but expressed in the notations of Ref.~\cite{menicucci2011graphical}. The physical origin of Eq.(\ref{eq:x_and_y}) resides in the requirement that the quadrature variances of the nullifiers Eq.(\ref{eq:quadratures_cluster}) tend to zero when squeezing tends to infinite. For this to hold, the coefficients multiplying the anti-squeezed quadratures terms must exactly cancel~\cite{vanLoock2007}. 
Consider a set of vacuum modes, with quadrature operators  $(\vec{x}^{(0)}, \vec{p}^{(0)})^T = (\hat{x}^{(0)}_1, ..., \hat{x}_{N}^{(0)},\hat{p}_{1}^{(0)}, ..., \hat{p}_{N}^{(0)})^T$. The transformations Eq.(\ref{eqtrasf-y0})
as well as the initial squeezing operation can be described by means of the symplectic matrices
\ba
\label{eq:boh2}
\left( \begin{array}{cccccccc}
\vec{x}^{\, '} \\
\vec{p}^{\, '} \\
\end{array} \right) &=&
\left( \begin{array}{cccccccc}
X_V & - Y_V \\
Y_V &  X_V \\
\end{array} \right)
\left(  \begin{array}{cccccccc}
K^{-\fr{1}{2}} & 0 \\
0 & K^{\fr{1}{2}}  \\
\end{array} \right)
\left( \begin{array}{cccccccc}
\vec{x}^{(0)} \\
\vec{p}^{(0)} \\
\end{array} \right) \nn \\
&=& 
\left( \begin{array}{cccccccc}
X_V K^{-\fr{1}{2}} \vec{x}^{(0)} - Y_V K^{\fr{1}{2}} \vec{p}^{(0)} \\
Y_V K^{-\fr{1}{2}} \vec{x}^{(0)} + X_V K^{\fr{1}{2}} \vec{p}^{(0)} \\
\end{array} \right),
\ea
where we have here explicited that 
$\left( \begin{array}{cccccccc}
\vec{x}^{\, \text{squ}} \\
\vec{p}^{\, \text{squ}} \\
\end{array} \right) =  \left(  \begin{array}{cccccccc}
K^{-\fr{1}{2}} & 0 \\
0 & K^{\fr{1}{2}}  \\
\end{array} \right)
\left( \begin{array}{cccccccc}
\vec{x}^{(0)} \\
\vec{p}^{(0)} \\
\end{array} \right)$, 
$K$ being the diagonal matrix representing the squeezing operation on each mode. In the simplest case of a uniform squeezing distribution $K = e^{-2 r} \mathcal{I}$ with $r$ real and positive, assuming that all the modes are $\hat p$-squeezed (the argument developed below is the same for a non-uniform squeezing distribution).
When building a cluster state with graph $V$, we want that the quadratures transformed according to Eq.(\ref{eq:boh2}) satisfy approximatively $\Delta^2 \hat \delta_i \rightarrow 0 \, \forall i$, with $\hat \delta_i$ given in Eq.(\ref{eq:quadratures_cluster}). In order to do so, we have to impose that the terms proportional to $e^{r}$ (i.e. to $K^{-\fr{1}{2}} $) are multiplied by zero exactly. 
We obtain
\ba
\label{eq:excesshot noise}
\vec{p}^{\, '} - V \vec{x}^{\, '} & &= (Y_V K^{-\fr{1}{2}} \vec{x}^{(0)} + X_V K^{\fr{1}{2}} \vec{p}^{(0)})  \\
   & &- V (X_V K^{-\fr{1}{2}} \vec{x}^{(0)} - Y_V K^{\fr{1}{2}} \vec{p}^{(0)} ) \rightarrow 0 \nn
\ea
which leads to 
\ba
\label{eq:magic_con}
&& (Y_V  - V X_V ) K^{-\fr{1}{2}} \vec{x}^{(0)} = 0 \hspace{0.3cm} \Rightarrow  \hspace{0.3cm} (Y_V  - V X_V ) = 0; \nn \\
&& \vspace{0.25cm} \nn  \\
&& (X_V + V Y_V ) K^{\fr{1}{2}} \vec{p}^{(0)} \rightarrow 0.
\ea
As mentioned, the first line in Eq.(\ref{eq:magic_con}) gives a physical meaning to the condition in Eq.(\ref{eq:x_and_y}). 
The remaining ``excess noise" quadratures in the second line provide variances
\be 
\label{eq:excess_noise}
\Delta^2 \hat \delta_i = \lqu (X_V + V Y_V) \mathcal{O} K \mathcal{O}^T (X_V + V Y_V)^T \rqu_{ii} 
\ee
which tend to zero in the infinite squeezing limit. 

\section{Measurement based quantum computation and calculation of the error on the result mode}
\label{app:B}

In this Appendix we provide all the details of the discussion in the main text, presenting optimized MBQC to reduce finite squeezing effects. 
With the choice of the beam splitter 
\be
U_{\text{BS}}^{\text{in},1} =\left(
\begin{array}{cccc}
 \frac{1}{\sqrt{2}} & \frac{i}{\sqrt{2}} & 0 & 0 \\
 \frac{i}{\sqrt{2}} & \frac{1}{\sqrt{2}} & 0 & 0 \\
 0 & 0 & 1 & 0 \\
 0 & 0 & 0 & 1
\end{array}
\right), 
\ee
the right quadratures to measure in order to realize teleportation of the input state onto the cluster are $\hat{x}_{\text{in}}$, $\hat{x}_1$.
For the Fourier transform discussed in the main text it turns out that a three-mode ancilla cluster is enough, and that the quadrature $\hat p_2$ should be measured on its second mode (as will be clear afterwards), in order to project the Fourier-transformed input state on the last mode of the cluster. This procedure is hence summarized in Eq.(\ref{eq:q_comp}), which expliciting the input-output modes reads
\be
\label{eq:boh7}
\left( \begin{array}{cccccccc}
\hat{a}_{\text{in}}^{'} \\
\hat{a}_1^{'} \\
\hat{a}_2^{'} \\
\hat{a}_3^{'} \\
\end{array} \right) =
D_{\text{meas}} U_{\text{BS}}^{\text{in},1} U_{\text{clu}}^{1,2,3} \left( \begin{array}{cccccccc}
\hat{a}_{\text{in}} \\
\hat{a}_1^{\text{squ}} \\
\hat{a}_2^{\text{squ}} \\
\hat{a}_3^{\text{squ}} \\
\end{array} \right) \equiv U_{\text{comp}} \left( \begin{array}{cccccccc}
\hat{a}_{\text{in}} \\
\hat{a}_1^{\text{squ}} \\
\hat{a}_2^{\text{squ}} \\
\hat{a}_3^{\text{squ}} \\
\end{array} \right).
\ee
\begin{figure} 
\begin{minipage}{.48\columnwidth}
\includegraphics*[width=\columnwidth]{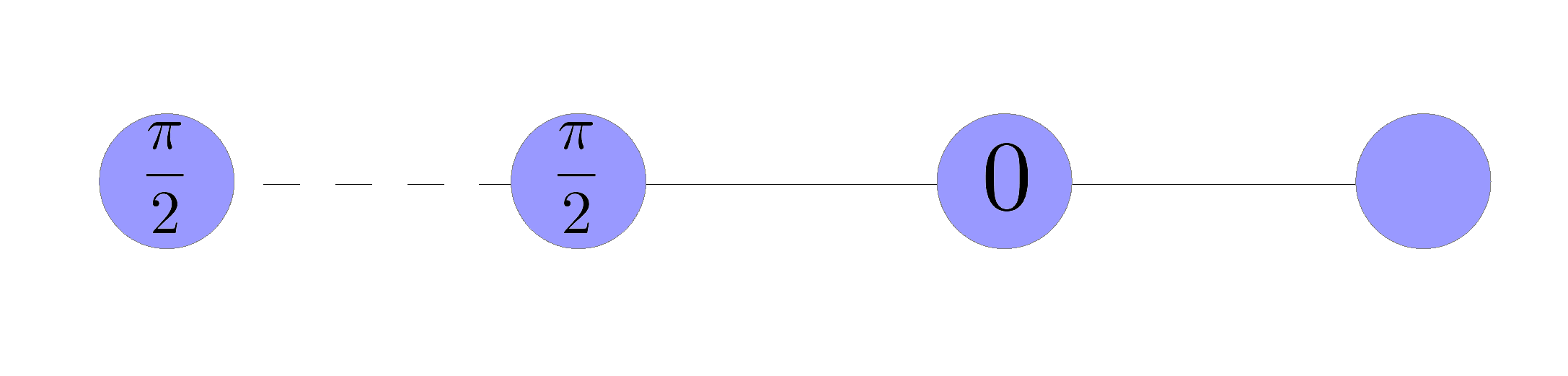}
\end{minipage}
\caption{A MBQC scheme where a Fourier transform is implemented on an input state by attaching it to a linear cluster state and by performing measurements on the cluster modes.}
\label{fig:cluster_TF}
\end{figure}

\subsection{Fixed realization of the cluster matrix}

For instance, let us consider first the case in which we chose as the matrix building the cluster state from independently squeezed modes Eq.(\ref{eq:sol_all}) the realization given in Ref.\cite{ferrini2013compact}. Then the matrices defining a Fourier transform quantum computation which appear in Eq.(\ref{eq:boh7}) are given by
{\small 
\ba
\label{eq:boh8}
\hspace{-0.4cm} D_{\text{meas}} &=& \text{diag} (i,i,1,1); \\
U_{\text{clu}}^{1,2,3} &=& 
\left(
\begin{array}{cccc}
 1 & 0 & 0 & 0 \\
 0 & 0 & -\sqrt{\frac{2}{3}} & -\frac{i}{\sqrt{3}} \\
 0 & -\frac{i}{\sqrt{2}} & -\frac{i}{\sqrt{6}} & -\frac{1}{\sqrt{3}} \\
 0 & -\frac{1}{\sqrt{2}} & \frac{1}{\sqrt{6}} & -\frac{i}{\sqrt{3}}
\end{array}
\right)  \nn
\ea}
from which
\be
U_{\text{comp}} =  D_{\text{meas}} U_{\text{BS}}^{\text{in},1} U_{\text{clu}}^{1,2,3} = \left(
\begin{array}{cccc}
 \frac{i}{\sqrt{2}} & 0 & \frac{1}{\sqrt{3}} & \frac{i}{\sqrt{6}} \\
 -\frac{1}{\sqrt{2}} & 0 & -\frac{i}{\sqrt{3}} & \frac{1}{\sqrt{6}} \\
 0 & -\frac{i}{\sqrt{2}} & -\frac{i}{\sqrt{6}} & -\frac{1}{\sqrt{3}} \\
 0 & -\frac{1}{\sqrt{2}} & \frac{1}{\sqrt{6}} & -\frac{i}{\sqrt{3}}
\end{array}
\right).
\ee
The squeezed input mode can be related to vacua modes as follows:
\be
\label{eq:squeezing4}
\left( \begin{array}{cccccccc}
\vec{x}^{\text{squ}} \\
\vec{p}^{\text{squ}} \\
\end{array} \right) =
\left(  \begin{array}{cccccccc}
K^{-\fr{1}{2}}& 0 \\
0 & K^{\fr{1}{2}} \\
\end{array} \right)
\left( \begin{array}{cccccccc}
\vec{x}^{(0)} \\
\vec{p}^{(0)} \\
\end{array} \right) %
\ee
with $K^{-\fr{1}{2}} = \text{diag}(1,e^{r_1},e^{r_2},e^{r_3})$ and $\vec{x}^{(0)} = ({x}_{\text{in}}^{(0)},{x}_1^{(0)},{x}_2^{(0)},{x}_3^{(0)})^T$.
By direct application of the transformation defined by Eq.(\ref{eq:boh7}) in the quadrature representation we obtain the quadratures of the transformed modes which are given by 
\ba
\label{eq:modi_vari5bis}
\left\{ \begin{array}{cccccccc}
\hat{x}'_{\text{in}}  &=& -\frac{\hat{p}_3^\text{squ}}{\sqrt{6}}-\frac{\hat{p}_{\text{in}}}{\sqrt{2}}+\frac{\hat{x}_2^\text{squ}}{\sqrt{3}}   \\
\hat{x}'_1  &=& \frac{\hat{p}_2^\text{squ}}{\sqrt{3}}+\frac{\hat{x}_3^\text{squ}}{\sqrt{6}}-\frac{\hat{x}_{\text{in}}}{\sqrt{2}}   \\
\hat{x}'_2  &=& \frac{\hat{p}_1^\text{squ}}{\sqrt{2}}+\frac{\hat{p}_2^\text{squ}}{\sqrt{6}}-\frac{\hat{x}_3^\text{squ}}{\sqrt{3}}   \\
\hat{x}'_3  &=&\frac{\hat{p}_3^\text{squ}}{\sqrt{3}}-\frac{\hat{x}_1^\text{squ}}{\sqrt{2}}+\frac{\hat{x}_2^\text{squ}}{\sqrt{6}}.
\end{array} \right.
\ea
and by 
\ba %
\label{eq:modi_vari5}
\left\{ \begin{array}{cccccccc}
\hat{p}'_{\text{in}}  &=&  \frac{\hat{p}_2^{\text{squ}}}{\sqrt{3}}+\frac{\hat{x}_3^{\text{squ}}}{\sqrt{6}}+\frac{\hat{x}_{\text{in}}}{\sqrt{2}} \hspace{0.25cm} a)   \\
\hat{p}'_1  &=& \frac{\hat{p}_3^{\text{squ}}}{\sqrt{6}}-\frac{\hat{p}_{\text{in}}}{\sqrt{2}}-\frac{\hat{x}_2^{\text{squ}} }{\sqrt{3}}  \hspace{0.25cm} b) \\
\hat{p}'_2  &=& -\frac{\hat{p}_3^{\text{squ}}}{\sqrt{3}}-\frac{\hat{x}_1^{\text{squ}}}{\sqrt{2}}-\frac{\hat{x}_2^{\text{squ}}}{\sqrt{6}}  \hspace{0.25cm} c) \\
\hat{p}'_3  &=& -\frac{\hat{p}_1^{\text{squ}}}{\sqrt{2}}+\frac{\hat{p}_2^{\text{squ}}}{\sqrt{6}}-\frac{\hat{x}_3^{\text{squ}}}{\sqrt{3}}. \hspace{0.25cm} d)
\end{array} \right. %
\ea
In this Heisenberg representation, the projective measurement of $\hat{p}'_{\text{in}},  \hat{p}'_1, \hat{p}'_2$ effectively results in replacing these operators by the corresponding measurement outcomes ${p}'_{\text{in}} $, $p'_1$ and ${p}'_2$ in Eq.(\ref{eq:modi_vari5})~\cite{Furusawa_Book}. Then, Eqs.(\ref{eq:modi_vari5}) and the last line in Eq.(\ref{eq:modi_vari5bis}) are solved for the output mode variables $\hat{p}'_3,\hat{x}'_3$, eliminating the anti-squeezed observables $\hat{x}_1^\text{squ},\hat{x}_2^\text{squ},\hat{x}_3^\text{squ}$. 
We obtain from (\ref{eq:modi_vari5}-a)  
\ba
\frac{\hat{x}_3^\text{squ}}{\sqrt{3}}  = \sqrt{2} {p}'_{\text{in}} - \sqrt{2} \frac{\hat{p}_2^\text{squ}}{\sqrt{3}} -  \hat{x}_{\text{in}}   \nn 
\ea
which  substituted in Eq. (\ref{eq:modi_vari5}-d) gives  
\be
\label{eq:seps5}
\hat{p}'_3  =  \hat{x}_{\text{in}}  - \fr{\hat{p}_1^\text{squ}}{\sqrt{2}} + 3 \fr{\hat{p}_2^\text{squ}}{\sqrt{6}} - \sqrt{2} {p}'_{\text{in}}.
\ee
From (\ref{eq:modi_vari5}-b)  
\ba
\hat{x}_2^\text{squ} = -\sqrt{3}{p}'_1 + \fr{1}{\sqrt{2}} \hat{p}_3^\text{squ} - \sqrt{\fr{3}{2}} \hat{p}_{\text{in}}
\ea
while from (\ref{eq:modi_vari5}-c)  we have
\be
\hat{x}_1^\text{squ} = -\sqrt{2} {p}'_2 -  \sqrt{\fr{2}{3}} \hat{p}_3^\text{squ} - \fr{1}{\sqrt{3}} \hat{x}_2^\text{squ} 
\ee
which  substituted in the equation for $\hat{x}'_3$ gives  
\be
\label{eq:seps6}
\hat{x}'_3  = -\hat{p}_{\text{in}} + p'_2 - \sqrt{2} p'_1 + \sqrt{3} \hat{p}_3^\text{squ}.
\ee
Identifying $\hat{x}_{\text{out}} = \hat{x}'_3$ and $\hat{p}_{\text{out}} = \hat{p}'_3$, from (\ref{eq:seps5}) and  (\ref{eq:seps6}) and making use of Eq.(\ref{eq:squeezing4}) we obtain 
\ba
\label{eq:teleport_result2}
\hat{x}_{\text{out}}  &=& \hat{x}'_3 = -\hat{p}_{\text{in}} + {p}'_2 - \sqrt{2} {p}'_1 +  \sqrt{3}  \hat{p}_3^{\text{squ}}  \\
\hat{p}_{\text{out}} &=& \hat{p}'_3 =  \hat{x}_{\text{in}}   - \sqrt{2} p'_{\text{in}} -  \fr{\hat{p}_1^{\text{squ}}}{\sqrt{2}} + 3 \fr{\hat{p}_2^{\text{squ}}}{\sqrt{6}}.  \nn
\ea
As mentioned in the main text, the result projected on the last mode of the cluster state (see Fig.~\ref{fig:cluster_TF}) is the desired Fourier transform of the input mode, plus some displacement which depends on the outcomes of the measurements performed on the previous modes (and which can be corrected by re-displacing back the last mode), as well as an undesired contribution due to the finite squeezing degree on the measured modes. The latter contributions eventually tend to zero when the squeezing degree goes to infinite in all the modes. Indeed the extra noise affecting the result associated with these undesired contributions is given by
\ba
\label{eq:extra_noise}
\hspace{-0.6cm} & \Delta^2 {\hat{x}_{\text{extra}} } = \Delta^2 \lqu  \sqrt{3} e^{-r_3} \hat{p}_3^{(0)} \rqu  = 3 e^{-2 r_3} \Delta^2_0 \nn \\
\hspace{-0.6cm} & \Delta^2 {\hat{p}_{\text{extra}} } = \Delta^2 \lqu  e^{-r_1} \fr{\hat{p}_1^{(0)}}{\sqrt{2}} + 3 e^{-r_2} \fr{\hat{p}_2^{(0)}}{\sqrt{6}} \rqu = \lt e^{- 2 r_1} + 3 e^{-2 r_2}  \rt \fr{\Delta^2_0}{2}. \nn
\ea
We can re-express the extra-noise contributions appearing in Eq.(\ref{eq:teleport_result2}) in terms of the cluster nullifiers. With the definition of the matrix leading to the cluster state expressed by Eq.(\ref{eq:boh8}) we have
\ba
\vec{a}^\text{clu} = U_{\text{clu}}^{1,2,3} \vec{a}^{\text{squ}}
\ea
(regarding only to the cluster modes) leading to
\ba
\label{eq:cluster-modes-x}
\left\{ \begin{array}{cccccccc}
\hat{x}_1^{\text{clu}}  &=& \frac{\hat{p}_3^\text{squ}}{\sqrt{3}}-\sqrt{\frac{2}{3}} \hat{x}_2^\text{squ}  \\
\hat{x}_2^{\text{clu}}  &=& \frac{\hat{p}_1^\text{squ}}{\sqrt{2}}+\frac{\hat{p}_2^\text{squ}}{\sqrt{6}}-\frac{\hat{x}_3^\text{squ}}{\sqrt{3}}\\
\hat{x}_3^{\text{clu}}  &=& \frac{\hat{p}_3^\text{squ}}{\sqrt{3}}-\frac{\hat{x}_1^\text{squ}}{\sqrt{2}}+\frac{\hat{x}_2^\text{squ}}{\sqrt{6}}
\end{array} \right.
\ea
and
\ba
\label{eq:cluster-modes-p}
\left\{ \begin{array}{cccccccc}
\hat{p}_1^{\text{clu}}  &=& -\sqrt{\frac{2}{3}} \hat{p}_2^\text{squ} - \frac{\hat{x}_3^\text{squ}}{\sqrt{3}}  \\
\hat{p}_2^{\text{clu}}  &=& -\frac{\hat{p}_3^\text{squ}}{\sqrt{3}}-\frac{\hat{x}_1^\text{squ}}{\sqrt{2}}-\frac{\hat{x}_2^\text{squ}}{\sqrt{6}}\\
\hat{p}_3^{\text{clu}}  &=& -\frac{\hat{p}_1^\text{squ}}{\sqrt{2}}+\frac{\hat{p}_2^\text{squ}}{\sqrt{6}}-\frac{\hat{x}_3^\text{squ}}{\sqrt{3}}.
\end{array} \right.
\ea
From Eqs.(\ref{eq:cluster-modes-x}) and (\ref{eq:cluster-modes-p}) one can compute the nullifiers 
\ba
\hat{\delta}_1 &=& \hat{p}_1^{\text{clu}} - \hat{x}_2^{\text{clu}}  = -\frac{\hat{p}_1^\text{squ} +\sqrt{3} \hat{p}_2^\text{squ}}{\sqrt{2}} \nn \\
\hat{\delta}_2 &=& \hat{p}_2^{\text{clu}}  - \hat{x}_1^{\text{clu}}  - \hat{x}_3^{\text{clu}}  = -\sqrt{3} \hat{p}_3^\text{squ}  \nn \\
\hat{\delta}_3 &=& \hat{p}_3^{\text{clu}}  - \hat{x}_2^{\text{clu}} = -\sqrt{2} \hat{p}_1^\text{squ}
\ea
It is then straightforward to re-express the terms of extra noise in Eqs.(\ref{eq:teleport_result2}) as
\ba
\sqrt{3} \hat{p}_3^\text{squ}  &=& - \hat{\delta}_2, \nn \\
- \fr{\hat{p}_1^\text{squ}}{\sqrt{2}} + 3 \fr{\hat{p}_2^\text{squ}}{\sqrt{6}} &=&  -\hat{\delta}_1 + \hat{\delta}_3
\ea
yielding to Eq.(\ref{eq:teleport_result4}) of the main text, where $\hat{\delta}_i$ are the nullifiers defined in Eq.(\ref{eq:quadratures_cluster}).

\subsection{Optimized realization of the cluster matrix}

We now run an evolutionary algorithm, which seeks to minimize Eq.(\ref{eq:fitness3}) over the angles $\vec{\theta}$, thereby reducing as much as possible the variances of the nullifiers providing extra noise in the result of the Fourier transform.
We obtain the optimized cluster matrix 
{\footnotesize  \ba
\label{eq:realization_best_U}
  & U_{\text{clu}}^{1,2,3} = \nn \\
  & \left(\cdot
\begin{array}{ccc}
 -9.8\cdot 10^{\text{-8}}+0.58 i & 0.71+\left(8.9\cdot 10^{\text{-8}}\right) i & 0.41-\left(1.5\cdot 10^{\text{-8}}\right) i \\
 0.58+\left(2.1\cdot 10^{\text{-8}}\right) i & 8.9\cdot 10^{\text{-8}}-\left(1.\cdot 10^{\text{-8}}\right) i & -1.5\cdot 10^{\text{-8}}+0.82 i \\
 1.2\cdot 10^{\text{-7}}+0.58 i & -0.71+\left(8.9\cdot 10^{\text{-8}}\right) i & 0.41-\left(1.5\cdot 10^{\text{-8}}\right) i
\end{array}
\right), \nn
\ea }
from which one can easily compute the cluster modes $\vec{a}^\text{clu} = U_{\text{clu}}^{1,2,3} \vec{a}^{\text{squ}}$. 
All the arguments presented above for the fixed realization of the cluster unitary matrix can be repeated; in particular, the output modes analogous to Eq.(\ref{eq:teleport_result2}) are given by 
\ba
\label{eq:teleport_result95}
\hat{x}_{\text{out}}   &=&  - \hat{p}_{\text{in}}  -1.7  \hat{p}_1^{\text{squ}}+\left(2.4\cdot 10^{-7}\right)  \hat{p}_2^{\text{squ}}+\left(5.5\cdot 10^{-8}\right)  \hat{p}_3^{\text{squ}} 
\nn \\ & & 
-1.4  {p}'_1 + {p}'_2   \\ 
\hat{p}_{\text{out}}   &=& \hat{x}_{\text{in}} + \left(2.\cdot 10^{-7}\right)  \hat{p}_1^{\text{squ}}+1.4  \hat{p}_2^{\text{squ}}-\left(1.3\cdot 10^{-8}\right)  \hat{p}_3^{\text{squ}} 
\nn  \\ & &
 -1.4 p'_{\text{in}} \nn
\ea
with $\hat p_i^{\text{squ}} = e^{- r_i} \hat{p}_i^{(0)}$.
Again, it is possible to recast the noise terms  appearing in Eq.(\ref{eq:teleport_result95}) in terms of the nullifiers, obtaining the same as in Eq.(\ref{eq:teleport_result4}). As shown in the main text, the value of the nullifier variances however depends on the specific realization of the matrix used, which allows choosing the best realization, yielding the lowest nullifier variances.

\bibliographystyle{apsrev}
\bibliography{multimode-BIB}

\end{document}